\documentclass[prl,aps,twocolumn,superscriptaddress]{revtex4-2}
\usepackage{amssymb}
\usepackage{epsfig}
\usepackage{graphicx}
\usepackage{amsmath}
\usepackage{array,color}
\usepackage{bm}
\usepackage{ragged2e}
\begin{document}

\title{Characterizing Superfluid Fraction and Topological Supersolid Phase Transitions via Nonequilibrium Dynamics}

\author{Biao Dong}
\address{MOE Key Laboratory for Nonequilibrium Synthesis and Modulation of Condensed Matter, Shaanxi Key Laboratory of Quantum Information and Quantum Optoelectronic Devices, School of Physics, Xi'an Jiaotong University, Xi'an 710049, China}
\author{Xiao-Fei Zhang}\email{xfzhang@sust.edu.cn}
\address{Department of Physics, Shaanxi University of Science and Technology, Xi'an 710021, China}
\author{Lin Zhuang}
\address{School of Physics, Sun Yat-Sen University, Guangzhou 510257, China}
\author{Wei Han}
\address{State Key Laboratory of Quantum Optics Technologies and Devices, Institute of Opto-Electronics, Collaborative Innovation Center of Extreme Optics, Shanxi University, Taiyuan, Shanxi 030006, China}
\author{Renyuan Liao}
\address{Fujian Provincial Key Laboratory for Quantum Manipulation and New Energy Materials, College of Physics and Energy, Fujian Normal University, Fuzhou 350117, China}
\address{Fujian Provincial Collaborative Innovation Center for Advanced High-Field Superconducting Materials and Engineering, Fuzhou 350117, China}
\author{Xue-Ying Yang}
\address{Key Laboratory of Time Reference and Applications, National Time Service Center, Chinese Academy of Sciences, Xi'an 710600, China}
\author{Wu-Ming Liu}\email{wmliu@iphy.ac.cn}
\address{Beijing National Laboratory for Condensed Matter Physics, Institute of Physics, Chinese Academy of Sciences, Beijing 100190, China}
\author{Yong-Chang Zhang}\email{zhangyc@xjtu.edu.cn}
\address{MOE Key Laboratory for Nonequilibrium Synthesis and Modulation of Condensed Matter, Shaanxi Key Laboratory of Quantum Information and Quantum Optoelectronic Devices, School of Physics, Xi'an Jiaotong University, Xi'an 710049, China}

\date{\today}

\begin{abstract}
Supersolids, simultaneously hosting superfluid and crystalline order, have been realized in quantum degenerate gases. However, quantitative probing of their superfluid fraction and topological properties remains a key challenge. Here, we explore the nonequilibrium dynamics of a spin-orbit coupled dipolar Bose-Einstein condensate subject to both driving and dissipation, and find an emergent oscillating current in the supersolid. The periodicity of the oscillating current is highly dependent on the superfluid fraction, which provides a promising pathway to measure supersolidity. Furthermore, we identify distinct topological supersolid states, with phase transitions detectable via nonequilibrium dynamics as well. This approach permits the exploration of a broad range of quantum many-body phenomena related to supersolidity, as well as topological properties, in dipolar condensates.
\end{abstract}

\maketitle

\begin{figure*}[tbp]
\centering
\includegraphics[width=1\textwidth]{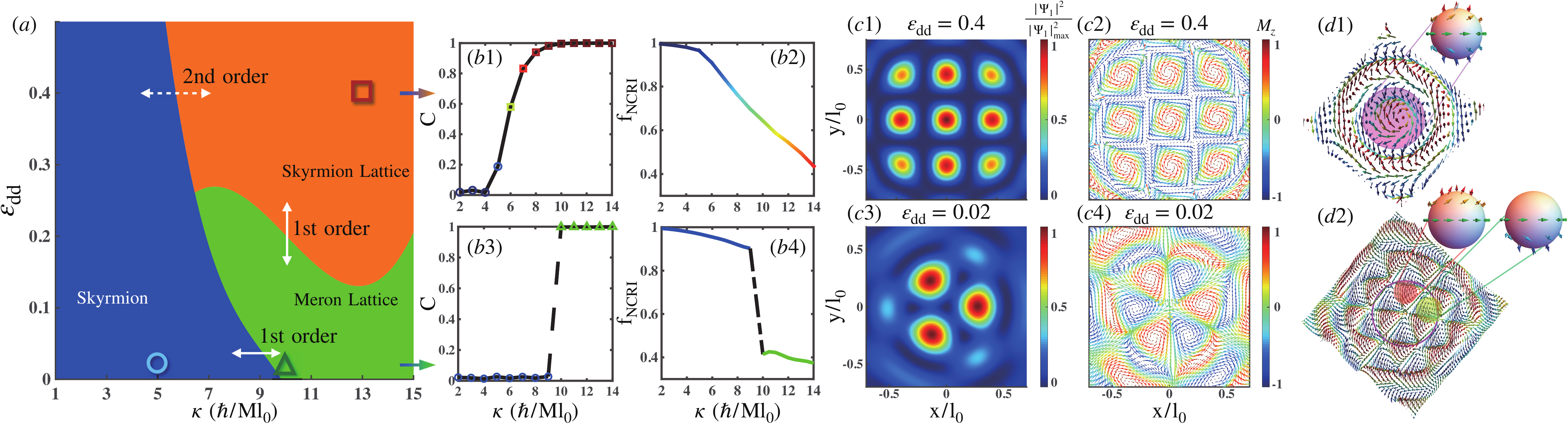}
\caption{(a) The phase diagram shows ground states of singly skyrmion (blue), skyrmion-type (orange), and meron-type (green) topological supersolids (TSSs) in a spin-$1/2$ dipolar condensate with SOC. Contact interactions are fixed at $g_{ij}N=g_{ii}N=gN=0.5$. The dipolar interaction ratio $\varepsilon_{\rm dd}=\frac{a_{\rm dd}}{a}$ and SOC strength $\kappa$ are varied. (b) Crystalline order (b1,b3) and superfluid fraction ($f_{\rm NCRI}$) (b2,b4) of the dipolar component evolve with $\kappa$, transiting from singly skyrmion to skyrmion-type (b1,b2) or meron-type (b3,b4) TSSs for strong ($\varepsilon_{\rm dd}=0.4$) or weak ($\varepsilon_{\rm dd}=0.02$) dipolar interactions, respectively. (c) Density profiles of the dipolar component (c1,c3) and spin textures (c2,c4) are shown for skyrmion-type TSS (c1,c2) at $\kappa=13$ and meron-type TSS (c3,c4) at $\kappa=10$. The color in (c2,c4) represents the $z$-component of spin density $M_z=\frac{1}{2}(|\Psi_1|^2-|\Psi_2|^2)/(|\Psi_1|^2+|\Psi_2|^2)$. (d) Spin textures of a skyrmion (d1) and meron pair (d2) map to 3D Bloch spheres, corresponding to the singly skyrmion (or one skyrmion of the skyrmion-type TSS) and meron-type TSS phases, respectively.}\label{F1}
\end{figure*}
\emph{Introduction}---The supersolidity of matter remains a pivotal goal in the exploration of novel states of matter~\cite{M. Boninsegni}. While initially pursued in solid $^{4}\textrm{He}$~\cite{A. F. Andreev,G. V. Chester,A. J. Leggett}, the field has been revitalized by the realization of supersolids in dipolar quantum gases~\cite{L. Tanzi2,M. Guo,F. Bottcher,L. Tanzi3,L. Chomaz,M. A. Norcia,J. Hertkorn,T. Bland}, following the achievement of quantum degeneracy in magnetic atoms~\cite{A. Griesmaier,M. Lu,K. Aikawa}. Within these ultracold atomic systems, exploring new states of matter characterized by topological configurations induced by synthetic gauge fields represents a fundamental area of research~\cite{L. Chomaz2}. These dipolar supersolids exhibit hallmark superfluid phenomena, including vanishing viscosity, quantized vortices, and nonclassical rotational inertia~\cite{S. Balibar,E. Casotti,L. Tanzi}, yet fundamental challenges persist, such as the direct measurement of superfluid fraction~\cite{G. Biagioni,L. Tanzi,G. Chauveau} and the engineering of topological orders.

Spin-orbit coupling (SOC) induced in ultracold systems provides unique opportunities to explore topological phenomena in supersolids~\cite{J. Leonard,J.-R. Li,S. Gopalakrishnan,Y. Deng,Q. Zhou,Y. Li,R. Liao,W. Han,K. T. Geier,K. T. Geier2}. Particularly, SOC-associated platforms enable spin textures inaccessible in single-component systems~\cite{H. Saito,J. Stenger} and facilitate chiral phase control relevant to topological quantum devices~\cite{A. Soumyanarayanan}. However, the effects of SOC on dipolar supersolids remains unclear.

In this Letter, we explore the ground-state properties of a two-component dipolar Bose-Einstein condensate (BEC) in the presence of spin-orbit coupling. The SOC and dipole-dipole interaction ($\varepsilon_{\text{dd}}$) strength~\cite{M. A. Baranov} act as competing energy scales in which the synthetic gauge field twists spins to generate topological spin textures that spontaneously break rotational symmetry and dipolar repulsion imposes spatial density modulation, creating a potential landscape wherein SOC-dominated effect locks spins into meron-antimeron pairs in a triangular lattice at low $\varepsilon_{\text{dd}}$, and dipolar interaction stabilizes skyrmions with square symmetry at high $\varepsilon_{\text{dd}}$. More profoundly, subjecting these equilibrium topological phases to $\mathcal{PT}$-symmetric driven-dissipation triggers oscillating currents whose frequency $\omega_d$ directly links to the superfluid fraction $f_{\text{NCRI}}$. This experimentally feasible scenario~\cite{K. Jimenez-Garcia,N. Q. Burdick,SM,C. Chin} establishes topological supersolids (TSSs) and provides a new approach to address the long-standing challenge of the superfluid-fraction measurement via nonequilibrium dynamics.

\emph{The spin-orbit coupled dipolar condensates}---We consider a two-component spin-orbit coupled BEC, where component $1$ consists of magnetic dipolar atoms and component $2$ is nonmagnetic~\cite{H. Saito,W. E. Shirley,B. Dong}. A magnetic field aligns the dipoles, and at zero temperature, the two-dimensional system is described by the macroscopic wave function $\mathbf{\Psi}=(\Psi_1,\Psi_2)$, governed by the Gross-Pitaevskii equation derived from the nondimensionalized mean-field energy~\cite{C. J. Pethick, L. Pitaevskii}
\begin{eqnarray}
\mathcal{E}[\Psi_1&,&\Psi_2]=\int d\mathbf{r}\biggl\{\sum_{j=1,2}
\bigg[ \Psi^\ast_{j}\left(-\frac{1}{2}\nabla^2+\frac{1}{2}r^2\right)\Psi_{j}\nonumber\\
&&+\frac{gN}{2}|\Psi_{j}|^4\bigg]+g_{12}N|\Psi_1|^2|\Psi_2|^2+\frac{g_{\rm dd}N}{2}\Phi_{\rm dd}|\Psi_1|^2 \nonumber\\
&&-i\kappa\left[\Psi^\ast_1(\partial_x-i\partial_y)\Psi_2+\Psi^\ast_2(\partial_x+i\partial_y)\Psi_1\right]\biggl\},\label{E1}
\end{eqnarray}
where $\int d\mathbf{r}(|\Psi_1|^2+|\Psi_2|^2)=1$. Here, $\Phi_{\rm dd}(\mathbf{r})=4\pi\mathcal{F}_{2D}^{-1}\left[\tilde{n}_1(\mathbf{k})F(\mathbf{k}/\sqrt{2})\right]$, with $\tilde{n}_1(\mathbf{k})=\mathcal{F}_{2D}[|\Psi_1(\mathbf{r})|^2]$ and $F(\mathbf{k})=2-3\sqrt{\pi}ke^{k^2}\mathrm{Erfc}(k)$, where $\mathcal{F}_{2D}$ is the two-dimensional Fourier transformation operator~\cite{U. R. Fischer,W. E. Shirley}. The SOC strength $\kappa$ and dipole-dipole interaction (DDI) strength $g_{\rm dd}=4\pi a_{\rm dd}/\sqrt{2\pi}l_z$ are controlled experimentally via Raman coupling in dipolar quantum gases (e.g., $^{52}\textrm{Cr}$) and magnetic Feshbach resonances, where $l_z=\sqrt{\hbar/M\omega_z}$ and $a_{\rm dd}=\frac{\mu_0\mu^2M}{12\pi\hbar^2}$~\cite{T. D. Lee2,T. D. Lee3,note2}. $\mu$ and $\mu_0$ are the dipole moment of the atom and the vacuum permeability, respectively. We explore the parameter space with $\kappa\in[1,15]$ and $\varepsilon_{\rm dd}\equiv\frac{a_{\rm dd}}{a}\in[0,0.5]$. The intra-component contact interactions are equal $g_{11}=g_{22}=g=4\pi a/\sqrt{2\pi}l_z$, with $a$ the $s$-wave scattering length. Energies and lengths are rescaled by $\hbar\omega_\perp$ and $l_0=\sqrt{\hbar/M\omega_\perp}$, respectively, where $M$ is the atomic mass and $\omega_\bot$ the radial trap frequency.

\emph{Topological supersolid phase transitions}---We numerically minimize the energy $\mathcal{E}[\Psi_1,\Psi_2]$ in Eq.~({\ref{E1}}) and explore ground-state phases by varying the dipolar-to-contact interaction ratio $\varepsilon_{\rm dd}=\frac{g_{\rm dd}}{g}$ and SOC strength~\cite{SM}. Without dipolar interactions ($\varepsilon_{\rm dd}=0$), strengthening SOC in a weakly interacting condensate can drive a phase transition from a half-quantum vortex to a hexagonal skyrmion lattice~\cite{H. Hu}. In the presence of DDI, the competition between DDI and SOC leads to richer phases with distinct topological and symmetry-breaking properties.

Fig.~{\ref{F1}}(a) summarizes our main results, showing the ground-state phase diagram as a function of SOC strength $\kappa$ and dipolar-to-contact interaction ratio $\varepsilon_{\rm dd}$. The diagram divides into two regions: a superfluid at weak SOC and a supersolid at strong SOC. For weak SOC, the condensate forms a single skyrmion without density modulation along the angle direction, characteristic of a superfluid. Beyond a critical $\kappa_c$, density modulation emerges, leading to crystallization into two types of topological supersolid states: meron and skyrmion lattices. The transition between these TSS states is controlled by $\varepsilon_{\rm dd}$. For small $\varepsilon_{\rm dd}$ (contact-dominated regime), the ground state exhibits a triangular density distribution and meron spin textures [Figs.~{\ref{F1}}(c3) and {\ref{F1}}(c4)], consistent with previous predictions for spin-orbit coupled BECs~\cite{H. Hu}. Surprisingly, we also find a skyrmion supersolid with a square density profile and skyrmion spin textures [Figs.~{\ref{F1}}(c1) and {\ref{F1}}(c2)]. Unlike single-component dipolar systems, which typically favor hexagonal structures, this square symmetry arises uniquely from the interplay of DDI and SOC.

For weak SOC ($\kappa<\kappa_c$), the wave function described by Eq.~({\ref{E1}}) spontaneously breaks U(1) symmetry, stabilizing a superfluid phase that hosts isolated skyrmions with topological charge $Q=\pm1$~\cite{N. Nagaosa}. Increasing $\kappa$ beyond $\kappa_c$ breaks rotational symmetry, stabilizing supersolids with triangular or square lattices depending on dipolar interaction strength [Figs.~{\ref{F1}}(a) and {\ref{F1}}(c)]. We observe a first-order transition from superfluid to meron supersolid and, unexpectedly, a second-order transition to skyrmion supersolid. Momentum-space densities [Fig.~{\ref{F3}}(a)] reveal the breaking of continuous rotational symmetry into three- and four-fold discrete rotational symmetries for meron and skyrmion supersolids. These phases are separated by a first-order transition at a critical dipolar interaction strength around $\varepsilon_{\rm dd}=0.2$ [Fig.~{\ref{F1}}(a)], forming a phase boundary across which the symmetry reconstruction occurs. The reciprocal lattice vector $|\mathbf{k}|$ increases (lattice constant $a_{\mathrm{sky}}$ decreases) with $\kappa$, constrained by $\sqrt{3}|\mathbf{k}|a_{\mathrm{sky}}/2=2\pi$ (triangular) and $|\mathbf{k}|a_{\mathrm{sky}}=2\pi$ (square). This behavior aligns with predictions for unconfined spin-orbit coupled condensates, where the roton minimum shifts to higher momentum as $\kappa$ increases~\cite{G. Juzeliunas,C.-J. Wu,R. Liao2}.

The superfluid-to-supersolid transition is accompanied by dramatic changes in magnetic properties. The singly skyrmion in the superfluid phase ($Q=\pm1$) is replaced by chiral lattices of skyrmions ($Q = 1$) or meron-antimeron pairs ($Q = \pm 1/2$) with distinct spin textures [Figs. 1(c2) and 1(c4)]. The topology of these textures is quantified by the topological charge $Q = \frac{1}{4\pi} \int d\mathbf{r} \, \hat{\mathbf{M}} \cdot (\frac{\partial \hat{\mathbf{M}}}{\partial x} \times \frac{\partial \hat{\mathbf{M}}}{\partial y})$, where $\hat{\mathbf{M}}=\mathbf{\Psi}^\dagger(\frac{1}{2}\hat{\bm{\sigma}})\mathbf{\Psi}/\mathbf{\Psi}^\dagger\mathbf{\Psi}$ is the local magnetization on the Bloch sphere~\cite{S. Heinze}. The square skyrmion lattice exhibits clockwise rotation ($Q=1$), while the triangular meron lattice features alternating meron-antimeron pairs ($Q=\pm1/2$) [Figs.~{\ref{F1}}(c4) and {\ref{F1}}(d2)]. These phases are further distinguished by their crystalline order and superfluid fraction, as detailed below.

\begin{figure}[tbp]
\centering
\includegraphics[width=0.45\textwidth]{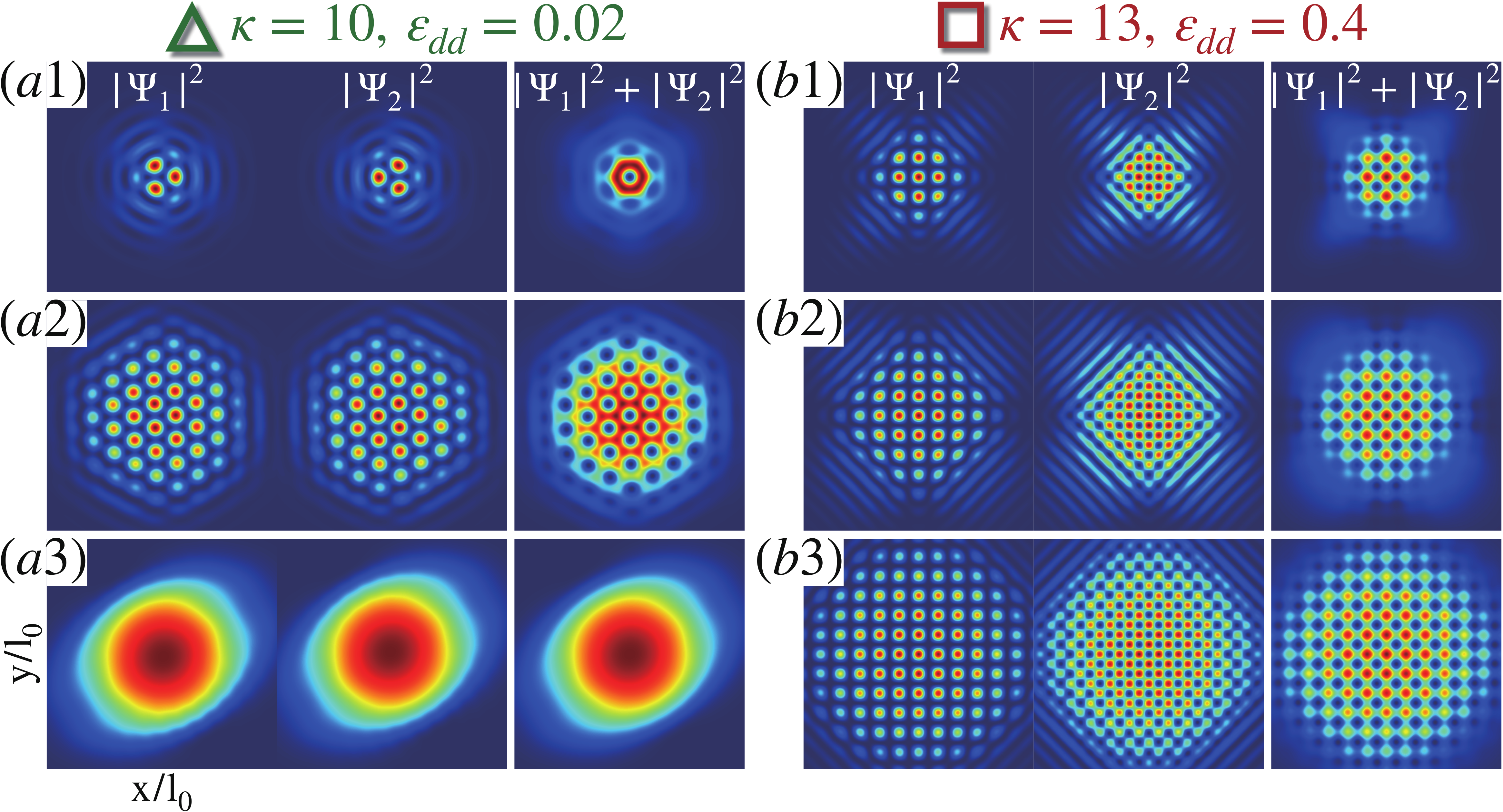}
\caption{Density profiles of (a) meron-type and (b) skyrmion-type TSSs are shown for varying intra- and inter-component contact interactions. The interaction strengths are (a1,b1) $g_{ii}N=g_{ij}N=0.5$, (a2,b2) $g_{ii}N=0.5,\ g_{ij}N=20$, and (a3,b3) $g_{ii}N=10,\ g_{ij}N=0.5$.}\label{F2}
\end{figure}

Tuning intra- and inter-component contact interactions enables precise control over density modulation and spin chirality, yielding supersolids with tailored droplet configurations and magnetic orders. Strong inter-component repulsion ($g_{ii}N=0.5,\ g_{ij}N=20$) drives density modulation, while intra-component interactions ($g_{ii}N=10,\ g_{ij}N=0.5$) stabilize a plane-wave phase with uniform spin alignment [Figs.~{\ref{F2}}(a2) and {\ref{F2}}(a3)]. For weak intra-component interactions ($g_{ii}N=0.5$), topological spin textures emerge at a critical balance: SOC promotes stripe phase, stabilizing a spin density wave, while dipolar interactions reinforce triangular (weak $\varepsilon_{\rm dd}$) or square (strong $\varepsilon_{\rm dd}$) lattice structures [Figs.~{\ref{F2}}(a2) and {\ref{F2}}(b2)]. Comparisons with contact-only systems~\cite{S. Sinha} highlight the essential role of dipolar repulsion in sustaining these hybrid phases. Under strong intra-component interactions ($g_{ii}N=10$), the system undergoes a transition from homogeneous superfluidity ($\kappa<\kappa'_c$) to topological supersolidity ($\kappa\geq\kappa'_c$), with strong SOC and DDI ($\kappa=13$, $\varepsilon_{\rm dd}=0.4$) stabilizing a square lattice [Figs.~{\ref{F2}}(a3) and {\ref{F2}}(b3)]. These results demonstrate the crucial role of interactions in structuring density and spin degrees of freedom, enabling the design of topological spin supersolids in spinor dipolar condensates.

\emph{Crystalline order and superfluid fraction}---To distinguish the crystalline properties of TSS from the superfluidity of the singly skyrmion phase, we employ two key observables: crystalline order, quantifying density modulation~\cite{Y.-C. Zhang,E. Poli}, and the nonclassical rotational inertia fraction (NCRIF), a measurement of supersolidity~\cite{L. Tanzi}. As shown in Figs.~{\ref{F1}}(b3) and {\ref{F1}}(b4), both crystalline order and NCRIF identify the transition from singly skyrmion superfluidity to meron supersolidity. Similarly, NCRIF exhibits discontinuous variation between meron and skyrmion supersolids, underscoring its sensitivity to changes in lattice symmetry and superfluid fraction.

The crystalline order $C=\frac{\rho_{\rm max}(\mathbf{k})-\rho_{\rm min}(\mathbf{k})}{\rho_{\rm max}(\mathbf{k})+\rho_{\rm min}(\mathbf{k})}$ quantifies density modulation in momentum space, where $\rho_{\rm max}(\mathbf{k})$ and $\rho_{\rm min}(\mathbf{k})$ are the maximum and minimum densities on the Rashba ring, respectively. For the meron lattice, $C$ jumps abruptly from $0.03$ to $1$ at $\kappa_c\approx9$ [Fig.~{\ref{F1}}(b3)], signaling a first-order transition from superfluid to topological meron supersolid.

To characterize the transition between TSSs, we evaluate the NCRIF, defined as
\begin{eqnarray}
f_{\rm NCRI}&=&\frac{\rho_s}{\rho_0}=\frac{1}{\langle\frac{1}{\rho_s(r)}\rangle\langle\rho(r)\rangle},\label{E3}
\end{eqnarray}
where $\rho(r)=\int d\varphi\rho(\mathbf{r},\varphi)/2\pi$ is the total density and $\rho_s(r)=2\pi/\int d\varphi\rho^{-1}(\mathbf{r},\varphi)$ is the superfluid density~\cite{A. J. Leggett,N. Henkel}. Here, $f_{\rm NCRI}=1$ ($0$) corresponds to a pure superfluid (solid), while $0<f_{\rm NCRI}<1$ indicates supersolidity. As shown in Fig.~{\ref{F1}}(b4), the NCRIF drops discontinuously from $0.9$ to $0.4$ at $\kappa_c\approx9$, confirming a first-order transition from superfluid to topological meron supersolid, consistent with the crystalline order results.

Figs.~{\ref{F1}}(b1) and {\ref{F1}}(b2) show the crystalline order and superfluid fraction across the transition from superfluid to skyrmion supersolid. Their continuous variation indicates a second-order transition, controlled by the SOC strength in the presence of strong dipolar interactions. As seen in the insets of Fig.~{\ref{F3}}(a), the modulated state comprises two pairs of orthogonal wavenumbers. Expanding energy of the system in terms of the modulation amplitude reveals a vanishing cubic term due to wavenumber orthogonality, explaining the second-order transition to a skyrmion square lattice~\cite{YCZ.Atoms}. Increasing SOC reduces the superfluid fraction below $1$, confirming crystallization into a supersolid. This second-order transition reflects gradual energy redistribution, consistent with the continuous evolution of the superfluid fraction.

\begin{figure}[tbp]
\centering
\includegraphics[width=0.45\textwidth]{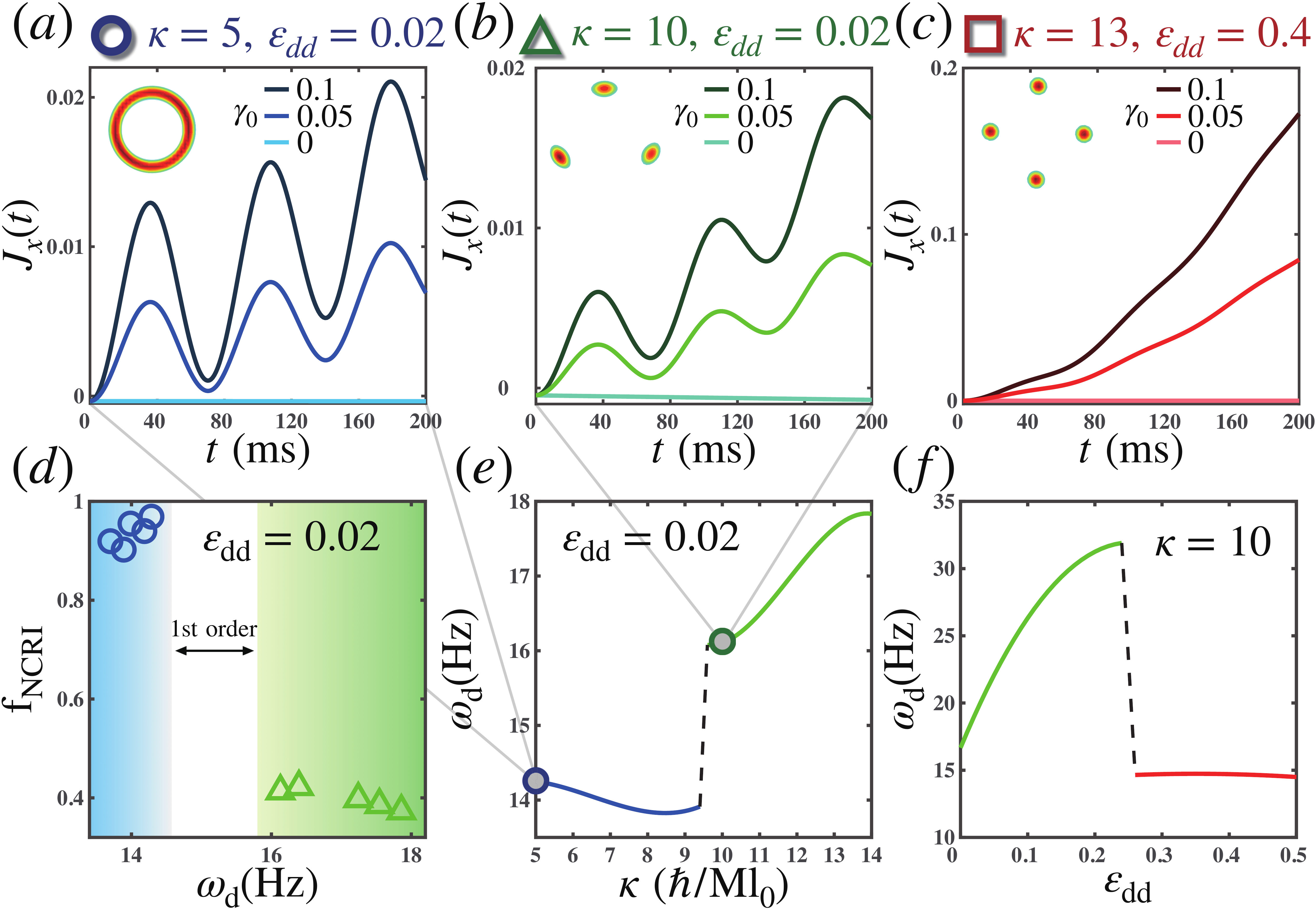}
\caption{Time evolution of the current $J_x(t)$ for (a) $\kappa = 5$, $\varepsilon_{dd}=0.02$; (b) $\kappa = 10$, $\varepsilon_{dd}=0.02$; and (c) $\kappa = 13$, $\varepsilon_{dd}=0.4$ [parameters from Fig.~\ref{F1}(a)]. Insets: Initial momentum-space profiles show symmetry breaking from (a) continuous rotational symmetry to (b) 3-fold (triangular) and (c) 4-fold (square) discrete rotational symmetry configurations. (d) Nonclassical rotational inertia fraction $f_{\text{NCRI}}$ versus driving frequency $\omega_d$ at $\varepsilon_{dd}=0.02$. (e) Discontinuous change in $\omega_d$ at the superfluid-to-supersolid phase transition (blue to green) versus $\kappa$ ($\varepsilon_{dd}=0.02$). (f) Discontinuous change in $\omega_d$ at the topological supersolid phase transition (green to red) versus $\varepsilon_{dd}$ ($\kappa=10$). Driving amplitude for (d)-(f) is $\gamma_0=0.02$.}\label{F3}
\end{figure}

\emph{Nonequilibrium probing via driven-oscillation}---To probe supersolidity-dissipation interplay, we introduce $\mathcal{PT}$-symmetric driving-dissipation~\cite{C. M. Bender,Y.-M. Hu,V. V. Konotop,C. M. Bender2,L. M. Sieberer,E. Zhao} into this quasi-2D system. The non-Hermitian Hamiltonian is~\cite{L. Pan}
\begin{eqnarray}
\mathcal{H}_{\mathrm{diss}}=-i\hbar\int d\mathbf{r}\mathbf{\gamma}(\mathbf{r})\sum_{j=1,2}|\Psi_{j}|^2, \label{E4}
\end{eqnarray}
where $\gamma(\mathbf{r})=\gamma_R(\mathbf{r})-\gamma_L(\mathbf{r})$ is the localized dissipation rate with $\gamma_l(\mathbf{r}) = \gamma_0 \exp\left[-\frac{(x-x_l)^2 + (y-y_l)^2}{2\sigma^2}\right]$ for $l = L, R$~\cite{V. A. Brazhnyi,G. Barontini,R. Labouvie}. Here, $\sigma=0.1$ and $\gamma_0$ are the driving (dissipation) width and amplitude, respectively, with gain and loss localized at $x_{R,L}=\pm0.8$ and $y_l=0$. This drives a current
\begin{eqnarray}
\mathbf{J}(t)
=-\frac{i}{2}\int{d\mathbf{r}\sum_{j=1,2}
{\left(\Psi_j^\ast\nabla\Psi_j-\Psi_j\nabla\Psi_j^\ast\right)}}. \label{E5}
\end{eqnarray}
linking equilibrium superfluid fraction to nonequilibrium dynamics (Fig.~{\ref{F3}}).

In the singly-skyrmion superfluid phase, undamped $J_x(t)$ oscillations [Fig.~{\ref{F3}}(a)] reflect global phase coherence~\cite{J. Williams}, yielding a near-unity superfluid fraction ($f_{\text{NCRI}}\approx 1$). In the meron supersolid phase [Fig.~{\ref{F3}}(b)], partial damping arises from triangular lattice rigidity and fragmented superflow, reducing $f_{\rm NCRI}$ to $0.4$. For the skyrmion supersolid [Fig.~{\ref{F3}}(c)], rapid suppression of $J_x(t)$ oscillations reveals square-lattice induced damping, with $f_{\rm NCRI}=0.5$. This demonstrates the damping depends not only on the superfluid fraction [Figs.~{\ref{F1}}(b4), {\ref{F3}}(a) and {\ref{F3}}(b)], but also on the lattice symmetry of the topological supersolids [Figs.~{\ref{F3}}(b) and {\ref{F3}}(c)]. These results highlight the role of oscillation damping in identifying topological supersolid phase transitions. Our open-system framework reveals how lattice symmetry and superfluid fraction, key signatures of supersolidity, govern driven-dissipative responses.

Discontinuous changes in the nonclassical rotational inertia fraction $f_{\rm NCRI}$ versus driven-oscillation frequency $\omega_d$ [Fig.~{\ref{F3}}(d)], and in $\omega_d$ versus SOC or dipolar interaction strengths [Figs.~{\ref{F3}}(e) or {\ref{F3}}(f)], reveal first-order superfluid-to-supersolid and topological supersolid phase transitions. For the superfluid-to-supersolid transition, the oscillation frequency jumps from $\omega_d\approx14$ to $\omega_d>16$ as SOC strength $\kappa$ increases [Fig.~{\ref{F3}}(e)], signaling a first-order transition from a singly-skyrmion superfluid to a meron-lattice supersolid. This provides a pathway to identify supersolidity by mapping the oscillation frequency to the superfluid fraction via driven nonequilibrium currents [Fig.~{\ref{F3}}(d)]. A similar discontinuity occurs for the topological supersolid transition when dipolar interaction strength $\varepsilon_{dd}$ increases [Fig.~{\ref{F3}}(f)]. These nonequilibrium results align with equilibrium measurements of crystalline order and superfluid fraction (Fig.~{\ref{F1}}). Figs.~{\ref{F3}}(d)-{\ref{F3}}(f) bridge equilibrium phase transitions and nonequilibrium dynamics, directly linking current responses to the superfluid fraction. The $f_{\rm NCRI}$-versus-$\omega_d$ relation quantifies supersolidity, analogous to recent advances in probing supersolid properties via nonequilibrium methods~\cite{G. Biagioni}.

On the experimental side, the system employs magnetic atoms ($^{52}\textrm{Cr}$, $^{164}\textrm{Dy}$, or $^{168}\textrm{Er}$) in a crossed optical dipole trap, powered by a laser at $1064$ $\mathrm{nm}$~\cite{A. Griesmaier}. The sublevels $|1\rangle=|m_J=-1\rangle$ or $|m_J=-2\rangle$ and $|2\rangle=|m_J=0\rangle$ are coupled via optical Raman dressing~\cite{M. Lecomte}, generating Rashba spin-orbit coupling $\hat{\mathcal{V}}_{\rm SO}=\tilde{\kappa}(\hat{p}_x\hat{\sigma}_x+\hat{p}_y\hat{\sigma}_y)$ with strength up to $\tilde{\kappa}=16.6\hbar/(Ml_0)$, where $l_0 = \sqrt{\hbar/M\omega_\perp}$ and $\omega_\perp = 2\pi \times 100$ Hz~\cite{Y. Deng,K. Jimenez-Garcia,N. Q. Burdick,SM}. Dipolar relaxation is suppressed via laser-induced quadratic Zeeman shifts~\cite{A. Griesmaier2,B. Pasquiou}, measured as $\Delta\omega_Z = 2\pi \times 5.6$ MHz between two sublevels in chromium~\cite{L. Santos2}. Contact interactions are tuned via Feshbach resonances at bias fields $B_0=590.5$ $\mathrm{G}$ (chromium), $B_0=7.1$ $\mathrm{G}$ (dysprosium), and $B_0=0.9$ $\mathrm{G}$ (erbium), while dipolar interactions are controlled by rotating magnetic field~\cite{S. Giovanazzi,Y. Tang}, achieving tunable relative strengths $\varepsilon_{\rm dd}$ ranging from $-12.08$ to $24.16$~\cite{SM}. The quasi-2D confinement is ensured by $V(\mathbf{r}) = \frac{1}{2}M(\omega_\perp^2\mathbf{r}^2+\omega_z^2z^2)$ with axial frequency $\omega_z=2\pi\times1$ $\mathrm{kHz}$ and aspect ratio $\lambda =\omega_z/\omega_\perp=10$~\cite{T. Lahaye}.

\emph{Conclusions}---We demonstrate the existence of topological supersolid phases in a spin-orbit coupled dipolar BEC, where crystalline order coexists with tunable spin textures. Crucially, nonequilibrium dynamics under $\mathcal{PT}$-symmetric driving-dissipation reveal an unexpected connection which directly quantifies the superfluid fraction by oscillation currents, establishing a new detection paradigm for supersolidity.

Our work bridges driven-dissipative dynamics and topological phase transitions in quantum many-body systems, establishing dipolar quantum gases as a versatile platform for integrating the supersolids with topological spin textures. Our results open avenues for exploring topological quantum matter and nonequilibrium dynamics in open systems. It resolves an interesting task and promotes exploring various topics, such as excitation spectra~\cite{P. B. Blakie,S. Lepoutre,L. M. Platt} and quantum fluctuations~\cite{B. Donelli}, in such system of future studies. All these studies could advance applications of topological supersolids in spintronics and topological quantum computing~\cite{A. P. Petrovic,S. L. Cornish}.

\emph{Acknowledgments}---This work was supported by National Key Research and Development Program of China (Grants No. 2021YFA1400900, No. 2021YFA0718300, No. 2021YFA1401700, and No. 2024YFF0726700), NSF of China (Grants No. 12104359, No. 12174461, No. 12234012, No. 12334012, and No. 52327808), Space Application System of China Manned Space Program, Shaanxi Academy of Fundamental Sciences (Mathematics, Physics) (Grant No. 22JSY036), Shaanxi Fundamental Science Research Project for Mathematics and Physics (Grant No. 23JSQ015), and the Fundamental Research Funds for the Central Universities (xpt012024033). Y.C.Z. acknowledges the support of Xi’an Jiaotong University through the “Young Top Talents Support Plan” and Basic Research Funding.


\begin{thebibliography}{99}
\bibitem{M. Boninsegni} M. Boninsegni and N. V. Prokof'ev, Rev. Mod. Phys. {\bf 84}, 759 (2012).

\bibitem{A. F. Andreev} A. F. Andreev and I. M. Lifshitz, Sov. Phys. JETP {\bf 29}, 1107 (1969).

\bibitem{G. V. Chester} G. V. Chester, Phys. Rev. A {\bf 2}, 256 (1970).

\bibitem{A. J. Leggett} A. J. Leggett, Phys. Rev. Lett. {\bf 25}, 1543 (1970).

\bibitem{L. Tanzi2} L. Tanzi, S. M. Roccuzzo, E. Lucioni, F. Fam\`{a}, A. Fioretti, C. Gabbanini, G. Modugno, A. Recati, and S. Stringari, Nature (London) {\bf 574}, 382 (2019).

\bibitem{M. Guo} M. Guo, F. B\"{o}ttcher, J. Hertkorn, J.-N. Schmidt, M. Wenzel, H. P. B\"{u}chler, T. Langen, and T. Pfau, Nature (London) {\bf 574}, 386 (2019).

\bibitem{F. Bottcher} F. B\"{o}ttcher, J.-N. Schmidt, M. Wenzel, J. Hertkorn, M. Guo, T. Langen, and T. Pfau, Phys. Rev. X {\bf 9}, 011051 (2019).

\bibitem{L. Tanzi3} L. Tanzi, E. Lucioni, F. Fam\`{a}, J. Catani, A. Fioretti, C. Gabbanini, R. N. Bisset, L. Santos, and G. Modugno, Phys. Rev. Lett. {\bf 122}, 130405 (2019).

\bibitem{L. Chomaz} L. Chomaz, D. Petter, P. Ilzh\"{o}fer, G. Natale, A. Trautmann, C. Politi, G. Durastante, R. M. W. van Bijnen, A. Patscheider, M. Sohmen, M. J. Mark, and F. Ferlaino, Phys. Rev. X {\bf 9}, 021012 (2019).

\bibitem{M. A. Norcia} M. A. Norcia, C. Politi, L. Klaus, E. Poli, M. Sohmen, M. J. Mark, R. N. Bisset, L. Santos, and F. Ferlaino, Nature (London) {\bf 596}, 357 (2021).

\bibitem{J. Hertkorn} J. Hertkorn, J.-N. Schmidt, M. Guo, F. B\"{o}ttcher, K. S. H. Ng, S. D. Graham, P. Uerlings, H. P. B\"{u}chler, T. Langen, M. Zwierlein, and T. Pfau, Phys. Rev. Lett. {\bf 127}, 155301 (2021).

\bibitem{T. Bland} T. Bland, E. Poli, C. Politi, L. Klaus, M. A. Norcia, F. Ferlaino, L. Santos, and R. N. Bisset, Phys. Rev. Lett. {\bf 128}, 195302 (2022).

\bibitem{A. Griesmaier} A. Griesmaier, J. Werner, S. Hensler, J. Stuhler, and T. Pfau, Phys. Rev. Lett. {\bf 94}, 160401 (2005).

\bibitem{M. Lu} M. Lu, N. Q. Burdick, S. H. Youn, and B. L. Lev, Phys. Rev. Lett. {\bf 107}, 190401 (2011).

\bibitem{K. Aikawa} K. Aikawa, A. Frisch, M. Mark, S. Baier, A. Rietzler, R. Grimm, and F. Ferlaino, Phys. Rev. Lett. {\bf 108}, 210401 (2012).

\bibitem{L. Chomaz2} L. Chomaz, I. Ferrier-Barbut, F. Ferlaino, B. Laburthe-Tolra, B. L. Lev, and T. Pfau, Rep. Prog. Phys. {\bf 86}, 026401 (2023).

\bibitem{S. Balibar} S. Balibar, Nature (London) {\bf 464}, 176 (2010).

\bibitem{E. Casotti} E. Casotti, E. Poli, L. Klaus, A. Litvinov, C. Ulm, C. Politi, M. J. Mark, T. Bland, and F. Ferlaino, Nature (London) {\bf 635}, 327 (2024).

\bibitem{L. Tanzi} L. Tanzi, J. G. Maloberti, G. Biagioni, A. Fioretti, G. Gabbanini, and G. Modugno, Science {\bf 371}, 1162 (2021).

\bibitem{G. Biagioni} G. Biagioni, N. Antolini, B. Donelli, L. Pezz\`{e}, A. Smerzi, M. Fattori, A. Fioretti, C. Gabbanini, M. Inguscio, L. Tanzi, and G. Modugno, Nature (London) {\bf 629}, 773 (2024).

\bibitem{G. Chauveau} G. Chauveau, C. Maury, F. Rabec, C. Heintze, G. Brochier, S. Nascimbene, J. Dalibard, J. Beugnon, S. M. Roccuzzo, and S. Stringari, Phys. Rev. Lett. {\bf 130}, 226003 (2023).

\bibitem{J. Leonard} J. L\'{e}onard, A. Morales, P. Zupancic, T. Esslinger, and T. Donner, Nature (London) {\bf 543}, 87 (2017).

\bibitem{J.-R. Li} J.-R. Li, J. Lee, W. Huang, S. Burchesky, B. Shteynas, F. \c{C}. Top, A. O. Jamison, and W. Ketterle, Nature (London) {\bf 543}, 91 (2017).

\bibitem{S. Gopalakrishnan} S. Gopalakrishnan, I. Martin, and E. A. Demler, Phys. Rev. Lett. {\bf 111}, 185304 (2013).

\bibitem{Y. Deng} Y. Deng, J. Cheng, H. Jing, C.-P. Sun, and S. Yi, Phys. Rev. Lett. {\bf 108}, 125301 (2012).

\bibitem{Q. Zhou} Q. Zhou and X. Cui, Phys. Rev. Lett. {\bf 110}, 140407 (2013).

\bibitem{Y. Li} Y. Li, Y. Liu, Z. Fan, W. Pang, S. Fu, and B. A. Malomed, Phys. Rev. A {\bf 95}, 063613 (2017).

\bibitem{R. Liao} R. Liao, Phys. Rev. Lett. {\bf 120}, 140403 (2018).

\bibitem{W. Han} W. Han, X.-F. Zhang, D.-S. Wang, H.-F. Jiang, W. Zhang, and S.-G. Zhang, Phys. Rev. Lett. {\bf 121}, 030404 (2018).

\bibitem{K. T. Geier} K. T. Geier, G. I. Martone, P. Hauke, and S. Stringari, Phys. Rev. Lett. {\bf 127}, 115301 (2021).

\bibitem{K. T. Geier2} K. T. Geier, G. I. Martone, P. Hauke, W. Ketterle, and S. Stringari, Phys. Rev. Lett. {\bf 130}, 156001 (2023).

\bibitem{J. Stenger} J. Stenger, S. Inouye, D. M. Stamper-Kurn, H.-J. Miesner, A. P. Chikkatur, and W. Ketterle, Nature (London) {\bf 396}, 345 (1998).

\bibitem{H. Saito} H. Saito, Y. Kawaguchi, and M. Ueda, Phys. Rev. Lett. {\bf 102}, 230403 (2009).

\bibitem{A. Soumyanarayanan} A. Soumyanarayanan, N. Reyren, A. Fert, and C. Panagopoulos, Nature (London) {\bf 539}, 509 (2016).

\bibitem{M. A. Baranov} M. A. Baranov, M. Dalmonte, G. Pupillo, and P. Zoller, Chem. Rev. {\bf 112}, 5012 (2012).

\bibitem{K. Jimenez-Garcia} K. Jim\'{e}nez-Garc\'{\i}a, L. J. LeBlanc, R. A. Williams, M. C. Beeler, C. Qu, M. Gong, C. Zhang, and I. B. Spielman, Phys. Rev. Lett. {\bf 114}, 125301 (2015).

\bibitem{N. Q. Burdick} N. Q. Burdick, Y. Tang, and B. L. Lev, Phys. Rev. X {\bf 6}, 031022 (2016).

\bibitem{SM} See Supplemental Material for more details on the experimental implementation of our proposal.

\bibitem{C. Chin} C. Chin, R. Grimm, P. S. Julienne, and E. Tiesinga, Rev. Mod. Phys. {\bf 82}, 1225 (2010).

\bibitem{W. E. Shirley} W. E. Shirley, B. M. Anderson, C. W. Clark, and R. M. Wilson, Phys. Rev. Lett. {\bf 113}, 165301 (2014).

\bibitem{B. Dong} B. Dong, Q. Sun, W.-M. Liu, A.-C. Ji, X.-F. Zhang, and S.-G. Zhang, Phys. Rev. A {\bf 96}, 013619 (2017).

\bibitem{C. J. Pethick} C. J. Pethick and H. Smith, \emph{Bose-Einstein Condensation in Dilute Gases} (Cambridge University Press, Cambridge, 2002).

\bibitem{L. Pitaevskii} L. Pitaevskii and S. Stringari, \emph{Bose-Einstein Condensation and Superfluidity} (Oxford University Press, Oxford, 2016).

\bibitem{U. R. Fischer} U. R. Fischer, Phys. Rev. A {\bf 73}, 031602(R) (2006).

\bibitem{T. D. Lee2} T. D. Lee and C. N. Yang, Phys. Rev. {\bf 105}, 1119 (1957).

\bibitem{T. D. Lee3} T. D. Lee, K. Huang, and C. N. Yang, Phys. Rev. {\bf 106}, 1135 (1957).

\bibitem{note2} We would like to point out that the Lee-Huang-Yang (LHY) correction~\cite{T. D. Lee2,T. D. Lee3} has been omitted in our theory Eq.~({\ref{E1}}), since the following discussion is focused on the weak dipolar interaction case. In such condition, the LHY effect is reasonably negligible which has been verified numerically.

\bibitem{H. Hu} H. Hu, B. Ramachandhran, H. Pu, and X.-J. Liu, Phys. Rev. Lett. {\bf 108}, 010402 (2012).

\bibitem{N. Nagaosa} N. Nagaosa and Y. Tokura, Nat. Nanotechnol. {\bf 8}, 899 (2013).

\bibitem{G. Juzeliunas} G. Juzeli\={u}nas, J. Ruseckas, and J. Dalibard, Phys. Rev. A {\bf 81}, 053403 (2010).

\bibitem{C.-J. Wu} C.-J. Wu, I. Mondragon-Shem, and X.-F. Zhou, Chin. Phys. Lett. {\bf 28}, 097102 (2011).

\bibitem{R. Liao2} R. Liao, O. Fialko, J. Brand, and U. Z\"{u}licke, Phys. Rev. A 92, 043633 (2015).

\bibitem{S. Heinze} S. Heinze, K. von Bergmann, M. Menzel, J. Brede, A. Kubetzka, R. Wiesendanger, G. Bihlmayer,
and S. Bl\"{u}gel, Nat. Phys. {\bf 7}, 713 (2011).

\bibitem{S. Sinha} S. Sinha, R. Nath, and L. Santos, Phys. Rev. Lett. {\bf 107}, 270401 (2011).

\bibitem{Y.-C. Zhang} Y.-C. Zhang, F. Maucher, and T. Pohl, Phys. Rev. Lett. {\bf 123}, 015301 (2019).

\bibitem{E. Poli} E. Poli, T. Bland, S. J. M. White, M. J. Mark, F. Ferlaino, S. Trabucco, and M. Mannarelli, Phys. Rev. Lett. {\bf 131}, 223401 (2023).

\bibitem{N. Henkel} N. Henkel, F. Cinti, P. Jain, G. Pupillo, and T. Pohl, Phys. Rev. Lett. {\bf 108}, 265301 (2012).

\bibitem{YCZ.Atoms} Y.-C. Zhang and F. Maucher, Atoms {\bf 11}, 102 (2023).

\bibitem{C. M. Bender} C. M. Bender and S. Boettcher, Phys. Rev. Lett. {\bf 80}, 5243 (1998).

\bibitem{Y.-M. Hu} Y.-M. Hu, H.-Y. Wang, Z. Wang, and F. Song, Phys. Rev. Lett. {\bf 132}, 050402 (2024).

\bibitem{V. V. Konotop} V. V. Konotop, J. Yang, and D. A. Zezyulin, Rev. Mod. Phys. {\bf 88}, 035002 (2016).

\bibitem{C. M. Bender2} C. M. Bender and D. W. Hook, Rev. Mod. Phys. {\bf 96}, 045002 (2024).

\bibitem{L. M. Sieberer} L. M. Sieberer, M. Buchhold, J. Marino, and S. Diehl, Rev. Mod. Phys. {\bf 97}, 025004 (2025).

\bibitem{E. Zhao} E. Zhao, Z. Wang, C. He, Ting Fung Jeffrey Poon, K. K. Pak, Y.- J. Liu, P. Ren, X.-J. Liu, and G.-B. Jo, Nature (London) {\bf 637}, 565 (2025).

\bibitem{L. Pan} L. Pan, X. Chen, Y. Chen, and H. Zhai, Nat. Phys. {\bf 16}, 767 (2020).

\bibitem{V. A. Brazhnyi} V. A. Brazhnyi, V. V. Konotop, V. M. P\'{e}rez-Garc\'{i}a, and H. Ott, Phys. Rev. Lett. {\bf 102}, 144101 (2009).

\bibitem{G. Barontini} G. Barontini, R. Labouvie, F. Stubenrauch, A. Vogler, V. Guarrera, and H. Ott, Phys. Rev. Lett. {\bf 110}, 035302 (2013).

\bibitem{R. Labouvie} R. Labouvie, B. Santra, S. Heun, and H. Ott, Phys. Rev. Lett. {\bf 116}, 235302 (2016).

\bibitem{J. Williams} J. Williams, R. Walser, J. Cooper, E. Cornell, and M. Holland, Phys. Rev. A {\bf 59}, R31 (1999).

\bibitem{M. Lecomte} M. Lecomte, A. Journeaux, J. Veschambre, J. Dalibard, and R. Lopes, Phys. Rev. Lett. {\bf 134}, 013402 (2025).

\bibitem{A. Griesmaier2} A. Griesmaier, J. Stuhler, and T. Pfau, Appl. Phys. B {\bf 82}, 211 (2006).

\bibitem{B. Pasquiou} B. Pasquiou, G. Bismut, Q. Beaufils, A. Crubellier, E. Mar\'{e}chal, P. Pedri, L. Vernac, O. Gorceix, and B. Laburthe-Tolra, Phys. Rev. A {\bf 81}, 042716 (2010).

\bibitem{L. Santos2} L. Santos, M. Fattori, J. Stuhler, and T. Pfau, Phys. Rev. A {\bf 75}, 053606 (2007).

\bibitem{S. Giovanazzi} S. Giovanazzi, A. G\"{o}rlitz, and T. Pfau, Phys. Rev. Lett. {\bf 89}, 130401 (2002).

\bibitem{Y. Tang} Y. Tang, W. Kao, K.-Y. Li, and B. L. Lev, Phys. Rev. Lett. {\bf 120}, 230401 (2018).

\bibitem{T. Lahaye} T. Lahaye, T. Koch, B. Fr\"{o}hlich, M. Fattori, J. Metz, A. Griesmaier, S. Giovanazzi, and T. Pfau, Nature (London) {\bf 448}, 672 (2007).

\bibitem{P. B. Blakie} P. B. Blakie, Phys. Rev. Lett. {\bf 134}, 013401 (2025).

\bibitem{S. Lepoutre} S. Lepoutre, A. Journeaux, J. Veschambre, J. Dalibard, and R. Lopes, Phys. Rev. A {\bf 111}, 033310 (2025).

\bibitem{L. M. Platt} L. M. Platt, D. Baillie, and P. B. Blakie, Phys. Rev. A {\bf 111}, 053305 (2025).

\bibitem{B. Donelli} B. Donelli, N. Antolini, G. Biagioni, M. Fattori, A. Fioretti, C. Gabbanini, M. Inguscio, L. Tanzi, G. Modugno, A. Smerzi, and L. Pezz\`{e}, arXiv:2501.17142.

\bibitem{A. P. Petrovic} A. P. Petrovi\'{c}, C. Psaroudaki, P. Fischer, M. Garst, and C. Panagopoulos, Rev. Mod. Phys. {\bf 97}, 031001 (2025).

\bibitem{S. L. Cornish} S. L. Cornish, M. R. Tarbutt, and K. R. A. Hazzard, Nat. Phys. {\bf 20}, 730 (2024).
\end{thebibliography}
\end{document}